\documentclass[aps,pra,twocolumn,showpacs,showkeys,superscriptaddress]{revtex4}

\usepackage{graphicx}
\usepackage{dcolumn}
\usepackage{bm}

\begin{document}


\title{Coherent pair production by photons in the 20-170 GeV energy
       range incident on crystals and birefringence}



\author{A.~Apyan}
\altaffiliation[Now at: ]{Northwestern University, Evanston, USA}
\affiliation{Yerevan Physics Institute, Yerevan, Armenia}

\author{R.O.~Avakian}
\affiliation{Yerevan Physics Institute, Yerevan, Armenia}

\author{B.~Badelek}
\affiliation{Uppsala University, Uppsala, Sweden}

\author{S.~Ballestrero}
\affiliation{INFN and University of Firenze, Firenze, Italy}

\author{C.~Biino}
\affiliation{INFN and University of Torino, Torino, Italy}
\affiliation{CERN, Geneva, Switzerland}

\author{I.~Birol}
\affiliation{Northwestern University, Evanston, USA}

\author{P.~Cenci}
\affiliation{INFN, Perugia, Italy}

\author{S.H.~Connell}
\affiliation{Schonland Research Centre - University of the Witwatersrand,
Johannesburg, South Africa}

\author{S.~Eichblatt}
\affiliation{Northwestern University, Evanston, USA}

\author{T.~Fonseca}
\affiliation{Northwestern University, Evanston, USA}

\author{A.~Freund}
\affiliation{ESRF, Grenoble, France}

\author{B.~Gorini}
\affiliation{CERN, Geneva, Switzerland}

\author{R.~Groess}
\affiliation{Schonland Research Centre - University of the Witwatersrand,
Johannesburg, South Africa}

\author{K.~Ispirian}
\affiliation{Yerevan Physics Institute, Yerevan, Armenia}

\author{T.J.~Ketel}
\affiliation{NIKHEF, Amsterdam, The Netherlands}

\author{Yu.V.~Kononets}
\affiliation{Kurchatov Institute, Moscow, Russia}

\author{A.~Lopez}
\affiliation{University of Santiago de Compostela, Santiago de Compostela,
Spain}

\author{A.~Mangiarotti}
\affiliation{INFN and University of Firenze, Firenze, Italy}

\author{B.~van~Rens}
\affiliation{NIKHEF, Amsterdam, The Netherlands}

\author{J.P.F.~Sellschop}
\altaffiliation[Deceased]{}
\affiliation{Schonland Research Centre - University of the Witwatersrand,
Johannesburg, South Africa}

\author{M.~Shieh}
\affiliation{Northwestern University, Evanston, USA}

\author{P.~Sona}
\affiliation{INFN and University of Firenze, Firenze, Italy}

\author{V.~Strakhovenko}
\affiliation{Institute of Nuclear Physics, Novosibirsk, Russia}

\author{E.~Uggerh{\o}j}
\thanks{Co-Spokeperson}
\affiliation{Institute for Storage Ring Facilities, University of Aarhus,
Denmark}

\author{U.I.~Uggerh{\o}j}
\affiliation{University of Aarhus, Aarhus, Denmark}

\author{G.~Unel}
\affiliation{Northwestern University, Evanston, USA}

\author{M.~Velasco}
\thanks{Co-Spokeperson}
\altaffiliation[Now at: ]{Northwestern University, Evanston, USA}
\affiliation{CERN, Geneva, Switzerland}

\author{Z.Z.~Vilakazi}
\altaffiliation[Now at: ]{University of Cape Town, Cape Town, South Africa}

\affiliation{Schonland Research Centre - University of the Witwatersrand,
Johannesburg, South Africa}

\author{O.~Wessely}
\affiliation{Uppsala University, Uppsala, Sweden}

\collaboration{The NA59 Collaboration}

\noaffiliation

\date{\today}

\begin{abstract}
The cross section for coherent pair production by linearly polarised
photons in the 20-170 GeV energy range was measured for photon aligned
incidence on ultra-high quality diamond and germanium crystals. The
theoretical description of coherent bremsstrahlung and coherent pair
production phenomena is an area of active theoretical debate and
development.  However, under our experimental conditions, the theory
predicted the combined cross section and polarisation experimental
observables very well indeed.  In macroscopic terms, our experiment
measured a birefringence effect in pair production in a crystal. This
study of this effect also constituted a measurement of the energy
dependent linear polarisation of photons produced by coherent
bremsstrahlung in aligned crystals. New technologies for manipulating high
energy photon beams can be realised based on an improved understanding of
QED phenomena at these energies. In particular, this experiment
demonstrates an efficient new polarimetry technique. The pair production
measurements were done using two independent methods simultaneously. The
more complex method using a magnet spectrometer showed that the simpler
method using a multiplicity detector was also viable.
\end{abstract}


\maketitle

\section{\label{sec:intro}Introduction}
This work focuses on polarisation phenomena in coherent bremsstrahlung
(CB) and coherent pair production (CPP) at high energies in oriented
single crystals. The CB and CPP theories are constructed in the framework
of the first Born approximation in the crystal
potential~\cite{cbdef1,cbdef2}.  These theories are well established and
experimentally investigated for relatively low energy (up to a few tens of
GeV) electrons and photons. The theoretical description of those phenomena
in oriented single crystals becomes very interesting at higher energies.
The processes have an strong angular and energy dependences and the
validity conditions of the Born approximation no longer hold at very high
energies and small incidence angles with respect to the crystal axes and
planes. The onset of this problem for the description of radiation
emission and pair production has the characteristic angle
$\theta_v=U_0/m$~\cite{baier}, where $U_0$ is the plane potential well
depth, $m$ is the electron rest mass and $\hbar=c=1$. The radiation and
pair production processes can be described by the CB and CPP theory for
the incidence angles with respect to the crystal axes/planes $\theta \gg
\theta_v$.  For angles $\theta \sim \theta_v$ and $\theta < \theta_v$ a
different approach, known as the quasi classical description is used. In
this approach the general theory of radiation and pair production are
developed based on the quasi classical operator
method~\cite{baier,strakh1}.

Historically, the pair conversion in a single crystals was proposed, and
later successfully used in the 1960s as a method to measure linear
polarisation for photons in the 1-6 GeV range~\cite{barbiellini}, It was
predicted theoretically and later verified experimentally~\cite{expver}
that the pair production cross section and the sensitivity to photon
polarisation increases with increasing energy. Therefore, at sufficiently
high photon energies, a new polarisation technique based on this effect
can be constructed, which will become competitive to other techniques,
such as pair production in amorphous media and photo nuclear methods.

In the experiment, the cross section for coherent pair production by
polarised photons incident on aligned germanium and diamond crystals was
measured, for different carefully selected crystallographic orientations.
This process can be effectively viewed as due to the imaginary part of the
refractive index, as it leads to an extinction of the photon beam. It
constitutes a birefringence phenomenon, as the imaginary part of the
refractive index will differ as a function of the angle between the plane
of polarisation of the photon beam and a specific crystallographic
orientation of the ``analyser'' crystal. A polarimeter was constructed by
measuring the energy dependent asymmetry with respect to the two most
distinct orientations of the analyser crystal with respect to pair
production.

The comparison to the data could validate the calculation of the energy
dependence of the cross section and the polarisation of photons produced
by coherent bremsstrahlung as well as the calculation of coherent pair
production for polarised photons incident on crystals of different
crystallographic orientations.

This paper starts with a brief discussion of the mechanism behind the
creation of the polarised photon beam and its Monte Carlo (MC)  
simulation. After presenting the motivations and the theory behind the
method used, the NA59 setup and analysis are discussed. The linear photon
polarisation measurement results using various analyser crystals are
followed by a comment on the possible applications of polarimetry with
aligned crystals.


\section{\label{production}Production of linearly polarised photons}

In the production of photon beams, single crystals can play an important
role by exploiting coherence and strong field effects that arise for
oriented incidence in the interaction of radiation and matter in
crystalline materials~\cite{pap1}. The Coherent Bremsstrahlung (CB) method
is a well established one for obtaining linearly polarised photons
starting from unpolarised electrons~\cite{cbdef1,cbdef2}. An electron
impinging on a crystal will interact coherently with the electric fields
of the atoms in aligned crystal planes. If the Laue condition is
satisfied, the bremsstrahlung photons will be emitted at specific energies
corresponding to the selected vectors of the reciprocal lattice. The
maximum polarisation and the maximum peak intensity occur at the same
photon energy, and this energy can be selected by choosing the orientation
of the lattice planes with respect to the incoming electron beam. This
property has been used previously to achieve photon beams with up to 70\%
linear polarization starting from 6\,GeV electrons~\cite{6gev}, and up to
60\% linear polarisation starting from 80~GeV electrons~\cite{omega}.

\begin{figure}[h]
\includegraphics[scale=0.48]{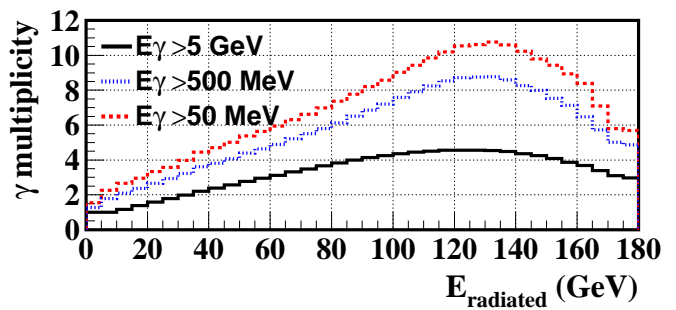}
\caption{\label{fig:multip} MC prediction for photon multiplicity vs total 
         radiated  energy  using different photon energy cut-off values.}
\end{figure}

The relative merits of different single crystals as CB radiators have been
investigated in the past~\cite{xtalprops}. The silicon crystal stands out
as a good choice due to its availability, ease of growth, and low mosaic
spread. The NA59 collaboration chose to use a 1.5~cm thick Si crystal to
achieve a relatively low photon multiplicity and reasonable photon
emission rate as seen in Fig.~\ref{fig:multip}. For an 178\,GeV electron
beam making an angle of 5\,mrad to the $<$001$>$ crystallographic axis and
about 70\,$\mu$rad from the (110)  plane, the resulting photon beam
polarisation spectrum was predicted to yield maximum polarisation of about
55\% in the vicinity of 70\,GeV.

For this choice of crystal orientation the incidence angles of electrons
and photons to the crystal plane become comparable with the radiation and
pair production characteristic angle $\theta_v$. In case of the $(110)$
plane of the silicon crystal, we find $\theta_v$=42 $\mu$rad. In fact part
of incident electron beam penetrates the crystal with angles both less and
greater than $\theta_v$, because of the angular divergence of the electron
beam.  In the theoretical simulations presented here, a Monte Carlo
approach was used to model the divergence of the electron and photon
beams, and the relevant theories (CB and CPP or the quasi classical
theory) are selected as appropriate for accurate and fast calculation.

It will be shown later that this approach has lead to a very good
agreement between the theoretical predications and the data.

\section{\label{polarimetry}Crystal polarimetry technique}

In this work, the photon polarisation is always expressed using the
Stoke's parametrisation with the Landau convention, where the total
elliptical polarisation is decomposed into two independent linear
components and a circular component. In mathematical terms, one writes:

\begin{widetext}
\begin{equation}
P_{\hbox {linear}}=\sqrt{\eta _{1}^{2}+\eta _{3}^{2}},
\quad \; P_{\hbox {circular}}=\sqrt{\eta _{2}^{2}},
\quad \; P_{\hbox {total}}=\sqrt{P_{\hbox {linear}}^{2}+P_{\hbox 
{circular}}^{2}} \quad .
\label{eq:pol-def}
\end{equation}
\end{widetext}

\begin{figure}[h]
\includegraphics[scale=0.43]{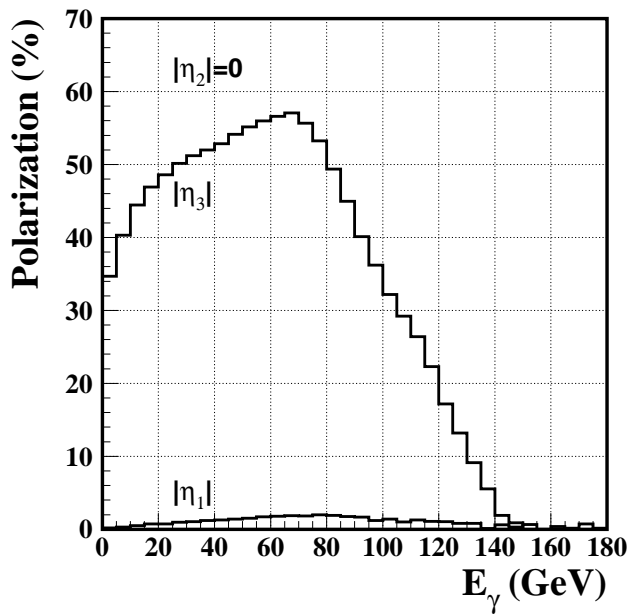}
\caption{\label{fig:pol-predict}Expected Polarisation for NA59 photon beam.}
\end{figure}

The radiator angular settings were chosen to have the total linear
polarisation from CB radiation purely along $\eta _{3}$. The NA59
collaboration thus made two distinct measurements, one to show that the
$\eta _{1}$ component of the polarisation was consistent with zero and
another to find the expected $\eta _{3}$ component of polarisation as
shown in Fig.~\ref{fig:pol-predict}.

The MC calculations used to obtain this prediction took into account the
divergence of the electron beam ($48\,\mu$rad horizontally and
33\,$\mu$rad vertically) and the 1\% uncertainty in its 178~GeV energy. To
optimise the processing time of the MC simulation, minimum energy cuts of
5~GeV for the electrons and 500 MeV for the photons were applied.  We
were, therefore, able to predict both the total radiated energy spectrum
and the energy spectrum of individual photons, as shown in
Fig.~\ref{fig:multip}.

\begin{figure}[ht]
\includegraphics[scale=0.43]{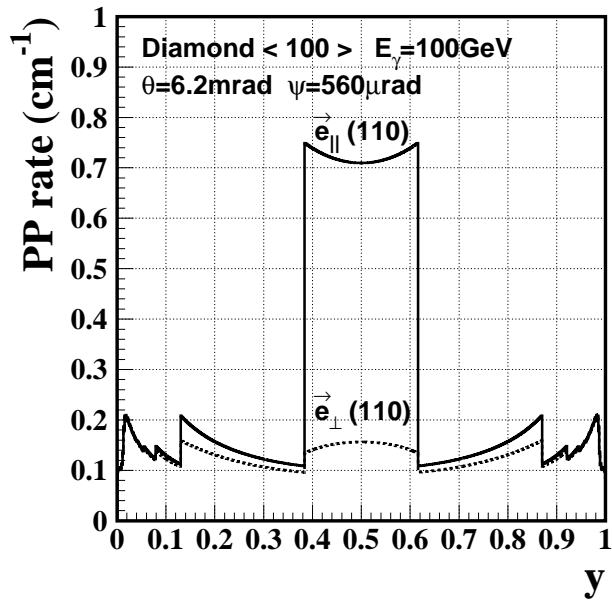}
\caption{\label{fig:ycut} Pair production rate vs the pair asymmetry, $y$, 
                as defined in the text.}
\end{figure}

\begin{figure*}[ht]
\includegraphics[scale=0.495]{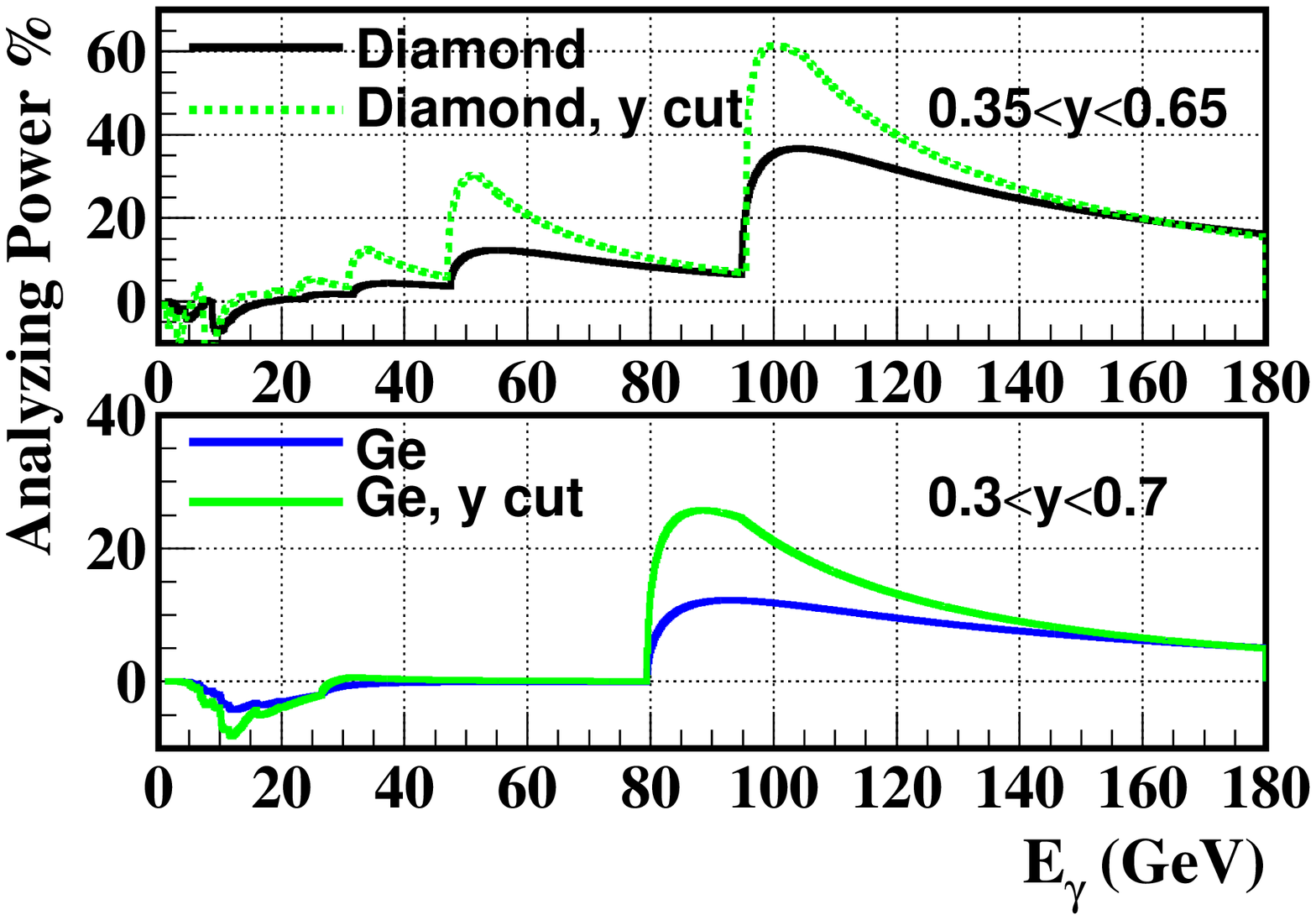}
\includegraphics[scale=0.495]{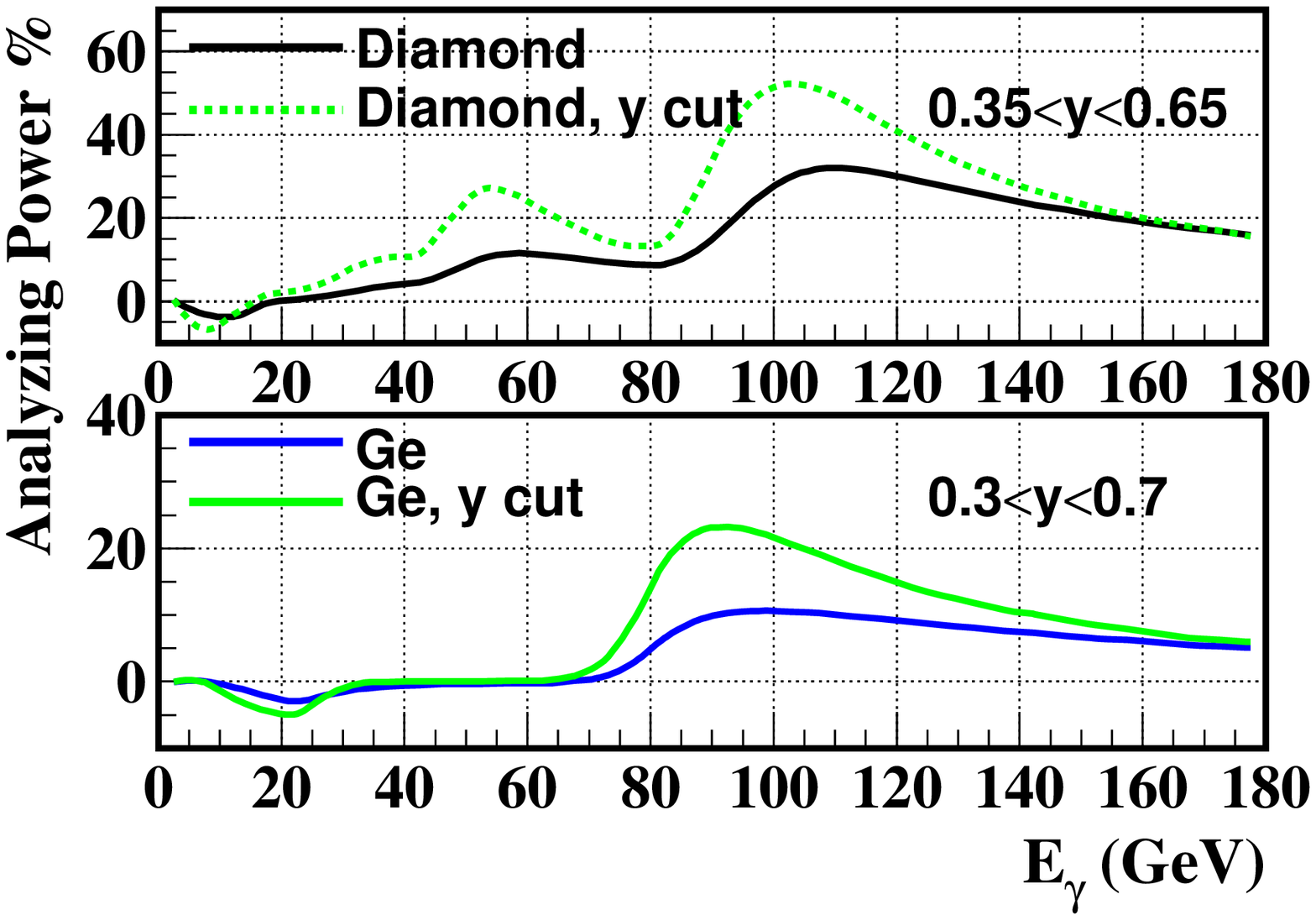}   
\caption{\label{fig:ap} Analysing power of different single crystals, for 
         an ideal $e^-$ beam without any angular divergences (left) and 
         for NA59 $e^-$ beam conditions (right).}
\end{figure*}

The polarisation dependence of the pair production cross section and the
birefringent properties of crystals are key elements of the photon
polarisation measurement.  The imaginary parts of the refraction indices
are related to the pair production cross section. This cross section is
sensitive to the relative angle between a crystal plane of a specific
symmetry and the plane of linear polarisation of the incident photon. In
essence, the two orthogonal directions where these two planes are either
parallel or perpendicular to each other yield the greatest difference in
pair production cross section. We therefore studied the pairs created in a
second aligned crystal, called the {\em analyser} crystal. In this study,
the experimentally relevant quantity is the asymmetry, $A$, between the
pair production cross sections, $\sigma$, of parallel and perpendicular
polarised photons, where the polarisation direction is measured with
respect to the (110) crystallographic plane of the analyser crystal. This
asymmetry is related to the linear photon polarisation, $P_{\rm l}$,
through the equation

\begin{equation}
A  \equiv   \frac{\sigma (\gamma _{\perp }\rightarrow e^{+}e^{-})-\sigma 
(\gamma _{\parallel }\rightarrow e^{+}e^{-})}{\sigma (\gamma _{\perp 
}\rightarrow e^{+}e^{-})+\sigma 
(\gamma _{\parallel }\rightarrow e^{+}e^{-})}
= R \times P_{\rm l}.
\label{eq:asy-def}
\end{equation}

Here $R$ is the so called ``analysing power'' of the second crystal. The
analysing power is in fact the asymmetry expected for the 100\% linearly
polarised photon beam. It will be seen that for the conditions of this
experiment, and using the theory described, this quantity can be reliably
computed using Monte Carlo simulations.  In this polarimetry method, the
crystal with the highest possible analysing power is preferred in order to
achieve a fast determination of the photon polarisation.

If one defines the ratio of the energy of one member of the pairs, $E^-$,
to the energy of the incoming photon, $E_\gamma$, as

\begin{equation}
y\equiv E^-/E_\gamma \quad ,
\end{equation}
then one may calculate the dependence of the pair production rate on this
ratio, $y$, as shown in Fig.~\ref{fig:ycut}. By comparing the rates for
the photon polarisation parallel (solid line) and perpendicular (dashed
line)  to the crystallographic plane, we observe that the largest
difference arises for \mbox{0.4 $\leq y \leq$ 0.6}.  Therefore the pair
production asymmetry may be maximised by selecting the subset of events
where the $e^{+}e^{-}$ pairs have similar energies. This method of
choosing the pairs to enhance the analysing power is called the
``quasi-symmetrical pair selection method''~\cite{ycut}. As a result of
such a cut, although the total number of events decreases, the relative
statistical error diminishes since it is inversely correlated with the
measured asymmetry. If the efficiencies of the pair events and beam
intensity normalisation events are assumed to be the same, then the cross
section measurement in equation~(\ref{eq:asy-def}) reduces to counting
these events separately. Denoting the number of pairs produced in
perpendicular and parallel cases by $p_{1}$ and $p_{2}$, and the number of
the normalisation events in each case by $n_{1}$ and $n_{2}$,
respectively, the measured asymmetry can be written as:

\begin{equation}
A=\frac{p_1/n_1 - p_2/n_2}{p_1/n_1 + p_2/n_2},
\label{eq:asy-meas}
\end{equation}
where $p$ and $n$ are acquired simultaneously and therefore correlated.

\subsection{Analyser Crystal Options}

The first analyser crystal used in the NA59 experiment was a germanium
(Ge) single crystal disk with a diameter of 3\,cm and a thickness of
1.0\,mm. The selected orientation with respect to the incident photon beam
represented a polar angle of 3.0\,mrad measured from the $<$110$>$ axis
and an azimuthal angle corresponding to incidence exactly on the (110)
plane. This configuration gave an analysing power peaking at 90~GeV, as
can be seen in Fig.~\ref{fig:ap}. From the same figure one can also see
that the quasi-symmetrical pair selection method delivers almost twice the
analysing power. The same single Ge crystal had also been used in the a
previous experiment, entitled NA43. The pair production properties of this
thickness of germanium crystal are therefore well known~\cite{Na43ge}.

The second analyser of the NA59 experiment was a multi-tile synthetic
diamond crystal target with an incident photon beam orientation with
respect to the crystal of 6.2\,mrad to the $<$001$>$ axis and
560\,$\mu$rad from the (110) plane.

\begin{figure}[h]
\includegraphics[scale=0.229]{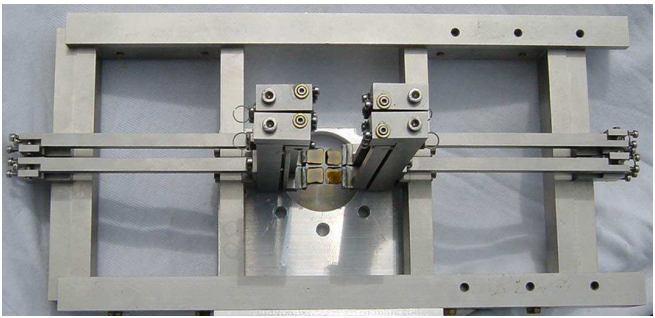}
\caption{\label{fig:diamonds} The diamond analyser target consists of
          synthetic diamond tiles and the aluminium holder frame.}
\end{figure}

The major advantage of using diamond in the analyser role are its high
pair yield, high analysing power (see Fig.~\ref{fig:ap}) and radiation
hardness. The photon beam dimensions of NA59 implied that one would need a
diamond with an area of about 20mm$\times$20mm. A crystal thickness of 4
mm was a fair compromise between requirements of the Figure of Merit for a
diamond analyser and the costs of the material. These requirements were
realised by developing a composite target comprising of four synthetic
diamonds of dimensions 8$\times$8$\times$4~mm$^3$ arranged in a square
lattice as seen in Fig.~\ref{fig:diamonds}. Selected areas of synthetic
diamonds grown under conditions of high pressure and temperature using the
``reconstitution'' technique~\cite{pap3} and other proprietary procedures
exhibit crystal structures superior to high quality natural samples. A
long term program of studying large synthetic single diamond crystals
using \mbox{X-ray} diffraction rocking curve widths, X-ray
topography~\cite{pap4,pap5}, cathodo-luminescence and indeed, experiments
with coherent bremsstrahlung and pair production in experiment NA43 on a
range of diamonds informed this procedure.

The four diamonds used in the composite analyser target were therefore
extracted from chosen regions of selected synthetic material yielding
regular tiles with optical surface finishes. The tiles had been polished
with faces corresponding to cubic directions with an accuracy of about 0.2
degrees using Laue back-reflection photographs. This pre-alignment fell
short of the requirements of the experiment. Each diamond tile should be
mutually aligned with its neighbours so that the $<$001$>$ axes normal to
the tile (approximately the beam direction)  corresponded within
5\,$\mu$rad. In addition, the mutual azimuthal alignment of the
crystallographic axes in the plane of the tile surfaces should be within
200\,$\mu$rad.

Accordingly, a mechanical system that featured three rotational degrees of
freedom for each diamond tile was designed. This rotation was effected by
mounting the diamond tiles on lever arms attached to a rigid frame using
spring loaded flexure hinges which could be actuated by very fine threaded
screws with significant mechanical advantage. Since the outer dimensions
of the whole four-crystal system were limited to 300 x 150 mm we had to
constrain the lengths of the lever arms to about 100~mm. The dynamic range
in angle space of the lever arms was necessarily limited, requiring the
diamond tiles to be pre-aligned in the adhesion mounting process. This was
achieved using a goniometer mounted vacuum tweezer to offer the tile to
the lever arm during fixing, under conditions of monitoring the
crystallographic orientation using an X-ray system. The adhesive used was
dental cement and the inter tile separation was 1 mm. The final accurate
mutual alignment was performed on a precise X-ray diffractometer at CNRS,
Grenoble/France using a well-collimated pencil beam. The whole alignment
system shown in Fig.~\ref{fig:diamonds} was mounted on a high precision
\mbox{XY-translation} table allowing each of the four crystals to be
illuminated with the pencil X-ray beam in turn. The slopes of the Bragg
peak at half maximum were used rather than the peak centre, as greater
sensitivity could be achieved in this way. This procedure was repeated
several times for all crystals to make sure that any cross-correlations
between the angular rotations were eliminated. Ultimately, all lever arms
were locked in position with dental cement to avoid any loss of the
adjustment by vibrations during transport from Grenoble to CERN.

This last procedure corresponded to the mutual alignment between the
elements of the composite target. The procedure was considered effective
and can form the basis of future aligned composite target systems.

An additional fine alignment is necessary in the orientation of radiator
crystal (Coherent Bremsstrahlung) and the analyser crystal (Asymmetry in
Pair Production for two orthogonal Analyser crystal orientations) with the
ideal particle in the beam envelope.

\begin{figure}[ht]
\includegraphics[scale=0.497]{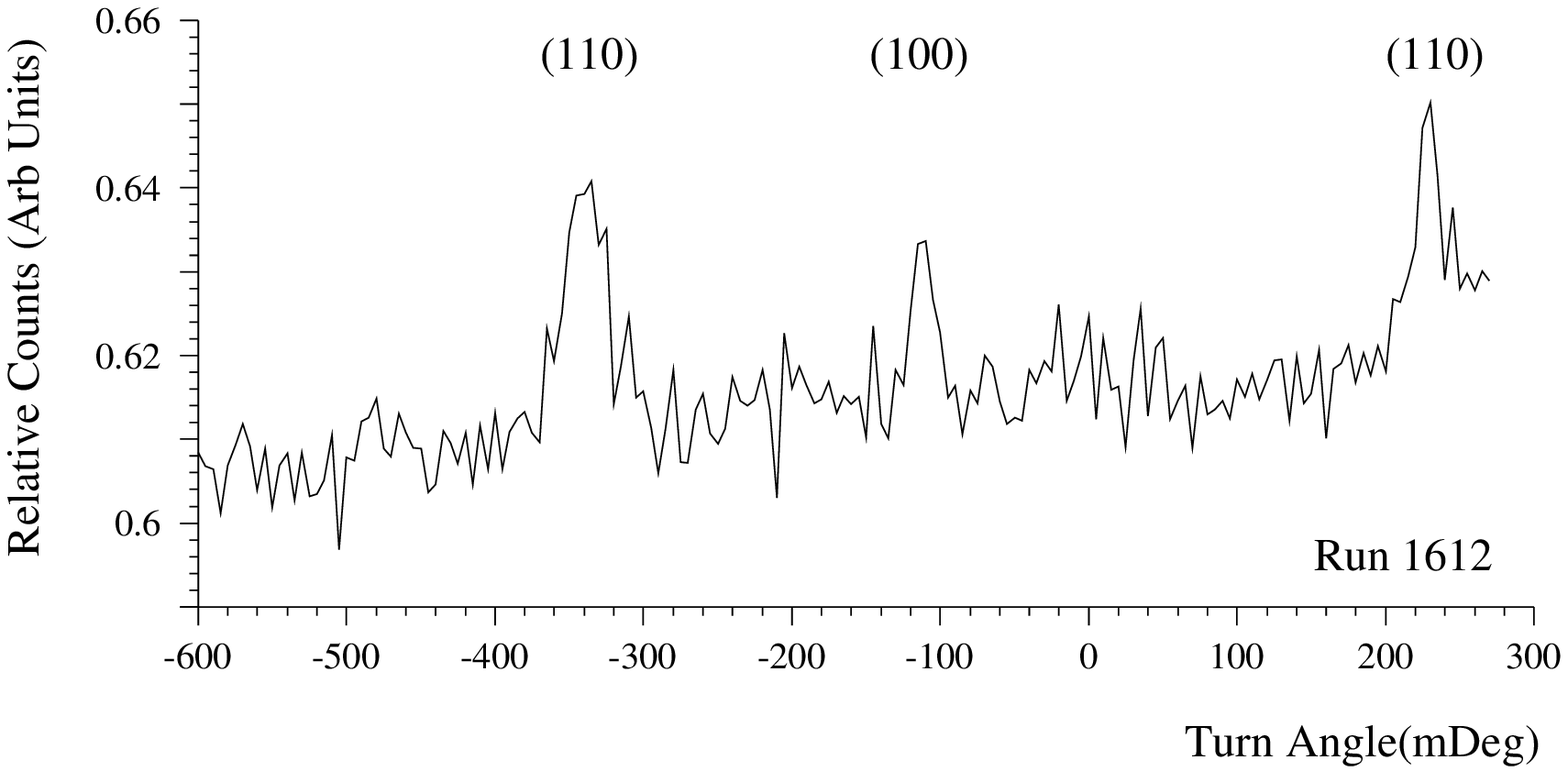}
\includegraphics[scale=0.491]{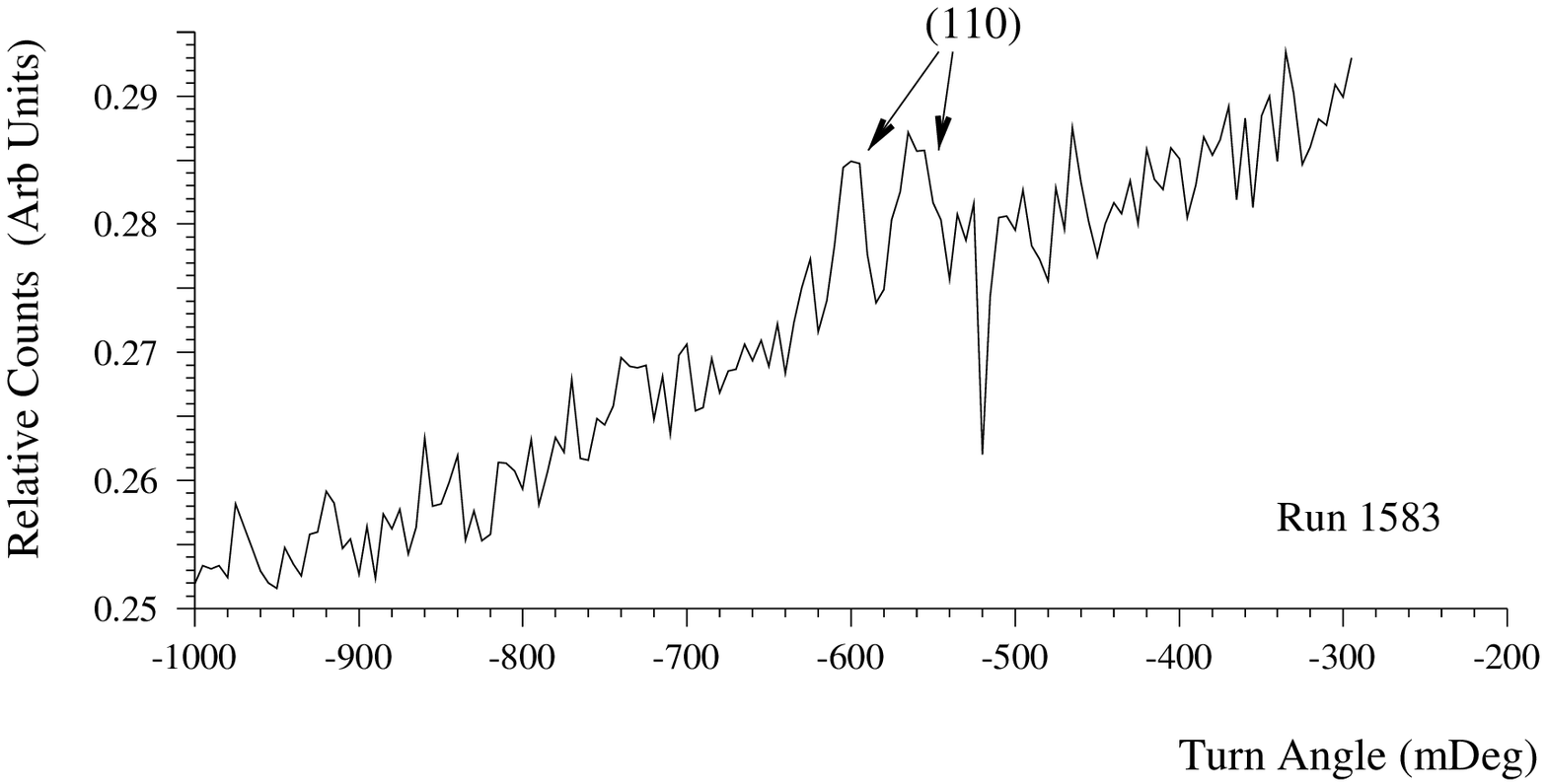}
\caption{\label{fig:1612}TOP: Scanning across the incident angle phase 
space between the beam and the crystal reveals coherent enhancements of 
the pair-production cross-section due to planar effects, as the planes 
are 
traversed in the scan. Stereograms generated from many orthogonal scans 
in 
the region around the axis allows the identification of the 
crystallographic planes.
BOTTOM: The mapping of the crystallographic planes revealed a misaligned
tile. The misaligned tile was identified in the offline analysis.}
\end{figure}

Once beam was available, the fine alignment was performed (and indeed
regularly controlled during the experiment). A narrow electron beam was
directed onto the crystal, and data was collected using the minimum bias
trigger. A scan of the incident angular phase space between the beam and
the crystal was performed by programming the motion of the crystal mounted
in the goniometers. The crystallographic axes and planes could be
identified as positions in this phase space where the coherent
enhancements (or reductions) of a radiation phenomenon in relation to the
corresponded incoherent cross-section occurred. This would be observed in
an appropriate detector.

The radiator crystal was therefore aligned exploiting the physics of
bremsstrahlung from the electron beam as observed in the Lead Glass
Calorimeter. On the other hand, the analyser crystal was aligned by
observing pair-production by the photon beam generated in the radiator
crystal as observed in the multiplicity counter.

Stereograms of the coherent enhancements were plotted in the incident
angle phase space. This allowed the planes around the axis to be
accurately identified and tracked. Usually, one would explore the region
around the axis, but off the axis, and then extrapolate the position of
the axis using the well understood and recognised behaviour of the
surrounding planes.

During the fine alignment process for the compound diamond target
(Fig.~\ref{fig:1612}), it was observed that one of the tiles of the
multi-crystal diamond analyser was accidentally misaligned. This lead to a
``doublet'' when one scanned across a plane. Analysis of the stereogram
indicated that one of the diamonds was out of alignment by 2.1 degrees
(Fig.~\ref{fig:1612}, bottom). This effect was incorporated in the offline
analysis where separate spectra for each diamond of the multi-diamond
target could be achieved. The offending tile was identified and excluded
from the analysis.

\section{Experiment and Analysis}
\subsection{The Setup}

\begin{figure*}[ht]
\includegraphics[scale=0.624]{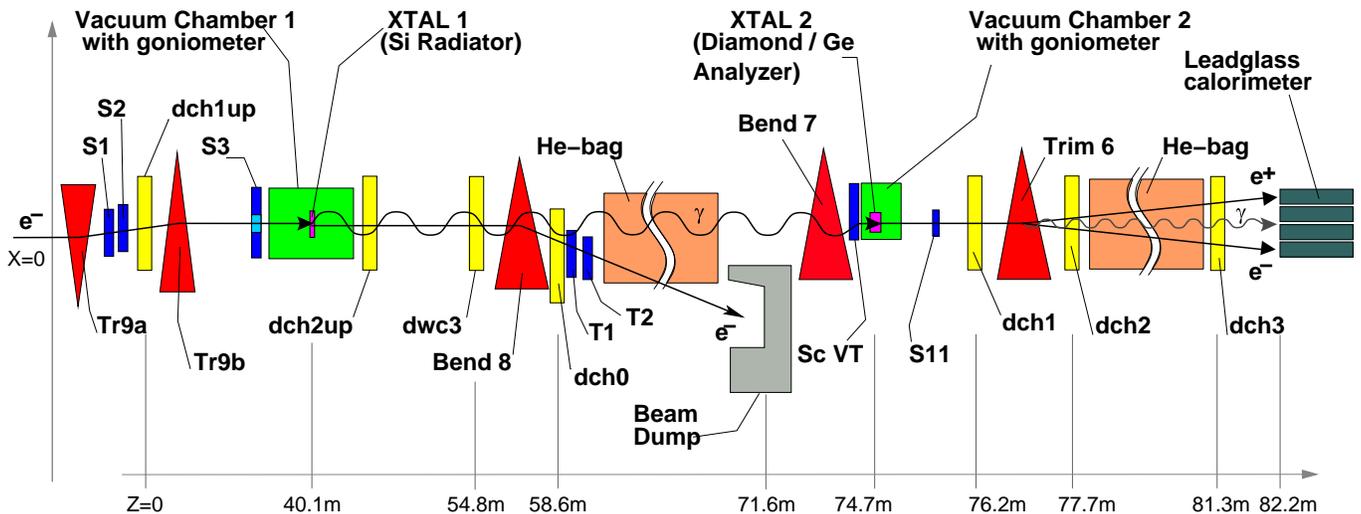}
\caption{\label{fig:exp-setup} The experimental setup.}
\end{figure*}

The experimental setup for the NA59 measurements in the year 2000 is shown
in Fig.~\ref{fig:exp-setup}. A 178~GeV beam of unpolarised electrons
from the CERN SPS accelerator was focused on the single crystal silicon
radiator (XTAL1). The crystal was of cylindrical shape with a 2.5~cm
radius and a 1.5~cm thickness, and it was aligned using a goniometer of
2~$\mu$rad precision to obtain CB radiation conditions. Upstream drift
chambers (dch1up-2up) allowed tracking of the incoming beam with an
angular precision of 4\,$\mu$rad.  The drift chambers had an active area
of 15$\times$15~cm$^2$ divided into six cells in both horizontal and
vertical planes. A double sense wire configuration removed the directional
hit ambiguity. The electron emerging from the radiator crystal was tagged
by two tracking chambers (dch2up and dwc3) to allow the measurement of its
multiple scattering angle inside the crystal. The dwc3 is a multi wire
proportional chamber~\cite{dwc} with an active area of about $10\times10$
cm and a resolution of 200~$\mu$m. A dipole magnet (Bend8) capable of a
maximum beam rigidity of 4.053 Tm and a special drift chamber (dch0) with
no active horizontal cells constituted the upstream spectrometer which
measured the energy of the electron, before sending it to the beam dump.

After passing a helium bag of length 9.65~m to reduce the multiple
scattering and background, the remaining photon beam impinged on the
analyser crystal aligned with a goniometer of 20~$\mu$rad precision. The
number of charged particles coming out of the analyser crystal was counted
both by a scintillator (S11) for fast triggering. The photons which did
not scatter or interact and the electron positron pairs created by the
interacting photons continued into a magnetic spectrometer. The dipole
analysis magnet (Trim6) of spectrometer was capable of a maximum beam
rigidity of 0.53 Tm. The tracking elements upstream of the magnet
consisted of one drift chamber (dch1) for the Ge analyser and two drift
chambers (dch05 and dch1) for the diamond analyser. There were two drift
chambers (dch2,dch3)  downstream of the magnet. The drift chambers
measured the horizontal and vertical positions of the passing charged
particles with a precision of 100\,$\mu$m yielding a spectrometer
resolution of ${\sigma}/{P^2}=0.12\%$. To measure the total radiated
energy, a 12-segment array lead-glass calorimeter of 24.6 radiation
lengths which had a resolution of ${\sigma}/{E}=11.5\%/\sqrt{E}$ was used.
The central segment of this lead-glass array was used to map and align the
crystals with an electron beam~\cite{expver}. A more detailed description
of the NA59 experimental apparatus is reported
elsewhere~\cite{mstheses1,mstheses2}.

Various plastic scintillators were used to calibrate the tracking chambers
and to define different physics triggers. The normalisation event trigger
($norm$) consisted of the signal logic combination
$S1{\cdot}S2{\cdot}\overline{\textrm S3}$ to ensure that an electron is
incident on the radiator crystal. The radiation event trigger ($rad$) was
defined as the signal logic combination
$norm{\cdot}(T1.or.T2){\cdot}\overline{VT}$ indicating that the incoming
electron has radiated and has been successfully taken out of the photon
section of the beam line. The pair event trigger ($pair$) was constructed
as the signal logic combination $rad{\cdot}S11$ to select the events for
which at least one $e^{+}e^{-}$ pair was created inside the analyser
crystal.

The NA59 data acquisition system consisted of personal computers running
the Linux operating system and using in-house developed software to access
the VME and CAMAC readout crates containing the digitisation modules. The
chamber signals were read out by VME TDC modules (Caen v767) with 1 ns
resolution. The scintillator and calorimeter signals were read out with
CAMAC ADC modules (LRS 2249)  with 0.3 pC resolution. The raw data were
then stored on DLT tapes for offline analysis.

\subsection{Analysis}

The first step in the offline analysis was the beam quality cuts, which
ensured the consistency of various trigger ratios and the initial beam
position and angles during data taking. Next, to facilitate comparison of
the experimental results with theoretical predictions, the angular
divergence of the electron beam was restricted to $\pm $3$\sigma$ from its
mean. Determination of the electron trajectory and its impact point on the
radiator were essential for fiducial volume requirements. The radiated
photons were taken to follow the direction of the initial electron. This
is accurate to $1/\gamma \approx 5\mu$rad for 100 GeV electrons. To
reconstruct the single photon energy in each event, only events where a
single electron positron pair was manifest in the spectrometer volume with
the pair energy being the same as the photon energy were selected. This
subset of pair events were further classified into families according to
the number of hits on the drift chambers of the spectrometer. In our
nomenclature, ``122 type'' events are clearly the cleanest ones with one
hit in the first upstream chamber, and two in both the second and third
downstream drift chambers. The resulting pair production vertex was
required to be in the fiducial volume of the analyser crystal. For the
case of the diamond analyser, the additional drift chamber on the
downstream side ensured a better vertex reconstruction. This in turn
allowed us to veto the inter-tile events as well as the ones coming from
the misaligned tile. Quality assessment of the pair search program was
performed by a GEANT based Monte Carlo (MC) program. This program
simulated the effects of the detector geometry to understand the precision
and efficiency of the reconstruction algorithm for each event family.

During the data taking, to obtain the parallel and perpendicular
configurations, the angular settings of the radiator crystal (hence the
direction of linear polarisation of the photon beam) were kept constant.
Only the analyser crystal was rotated in a rolling motion around its
symmetry axis. Therefore to measure the magnitude of the $\eta _{3}$
($\eta _{1}$) component of the polarisation, analyser orientations
separated by $\pi /2$ starting from 0 ($\pi /4$) were compared. To reduce
the systematic errors, (especially in the case of the Ge crystal where the
analysing power is smaller), all relevant angles on the analyser crystal
were utilised for polarisation measurements, as presented in Table
\ref{tab:data-sets}. Other sources of systematic errors were the
uncertainty in the crystal angles, the photon tagging and the pair
reconstruction efficiencies obtained from MC studies.

\begin{table}[h]
\caption{\label{tab:data-sets} Different material and angular settings for 
      the analyser crystal used to measure the linear polarisation 
      components.}
\renewcommand{\extracolsep}{}
\begin{tabular}{|c||c|c|}
\hline
\textbf{analyser orientation }&
\textbf{\ \ analyser\ \ }&
\textbf{measured polariza-}\\
\textbf{(roll wrt radiator)}&
 \textbf{type}&
\textbf{tion component}\\
\hline
\hline
$0,\frac{\pi }{2},\pi ,\frac{3\pi }{2}$& Ge& $\eta _{3}$\\ \hline
$\frac{\pi }{4},\frac{3\pi }{4},\frac{5\pi }{4},\frac{7\pi }{4}$& Ge&
$\eta _{1}$\\ \hline
$0,\frac{\pi }{2}$& Diamond& $\eta _{3}$\\ \hline
\end{tabular}
\end{table}

\subsection{CB Validation}

\begin{figure}[h]
\includegraphics[scale=0.468]{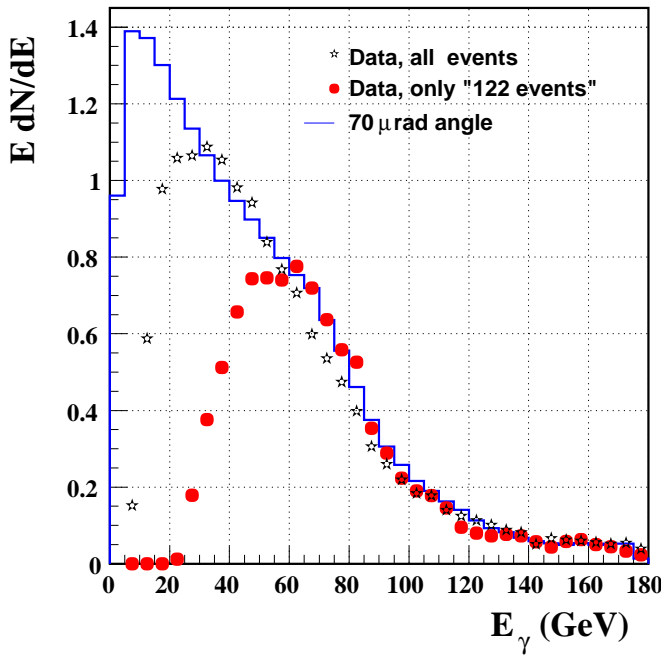}
\caption{\label{fig:single-photons} MC predictions for the single photon 
   spectrum,  compared with data using all events (stars) and only `122 
   family' events (circles).}
\end{figure}

The angular settings of the radiator crystal were inferred from the data.
The single photon intensity spectrum presented in
Fig.~\ref{fig:single-photons} contains two different event selections
superimposed on the MC prediction. The geometrical acceptance of the
spectrometer for these events has a high threshold of 30~GeV, as seen from
Fig.~\ref{fig:single-photons}. 

\begin{figure}[ht]
\includegraphics[scale=0.459]{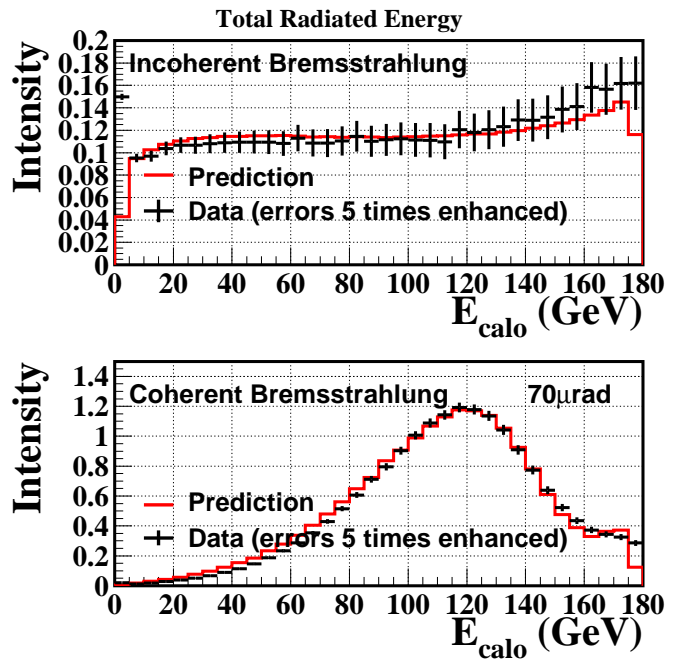}
\caption{\label{fig:pileup} Total energy radiated for Incoherent (top) and 
Coherent (bottom) bremsstrahlung radiation. Note the sensitivity of the 
cross section to small changes in the angular setting of the crystal. The 
statistical errors on the data are enhanced by a factor of five to 
increase visibility.}
\end{figure}

\begin{figure}[ht]
\includegraphics[scale=0.479]{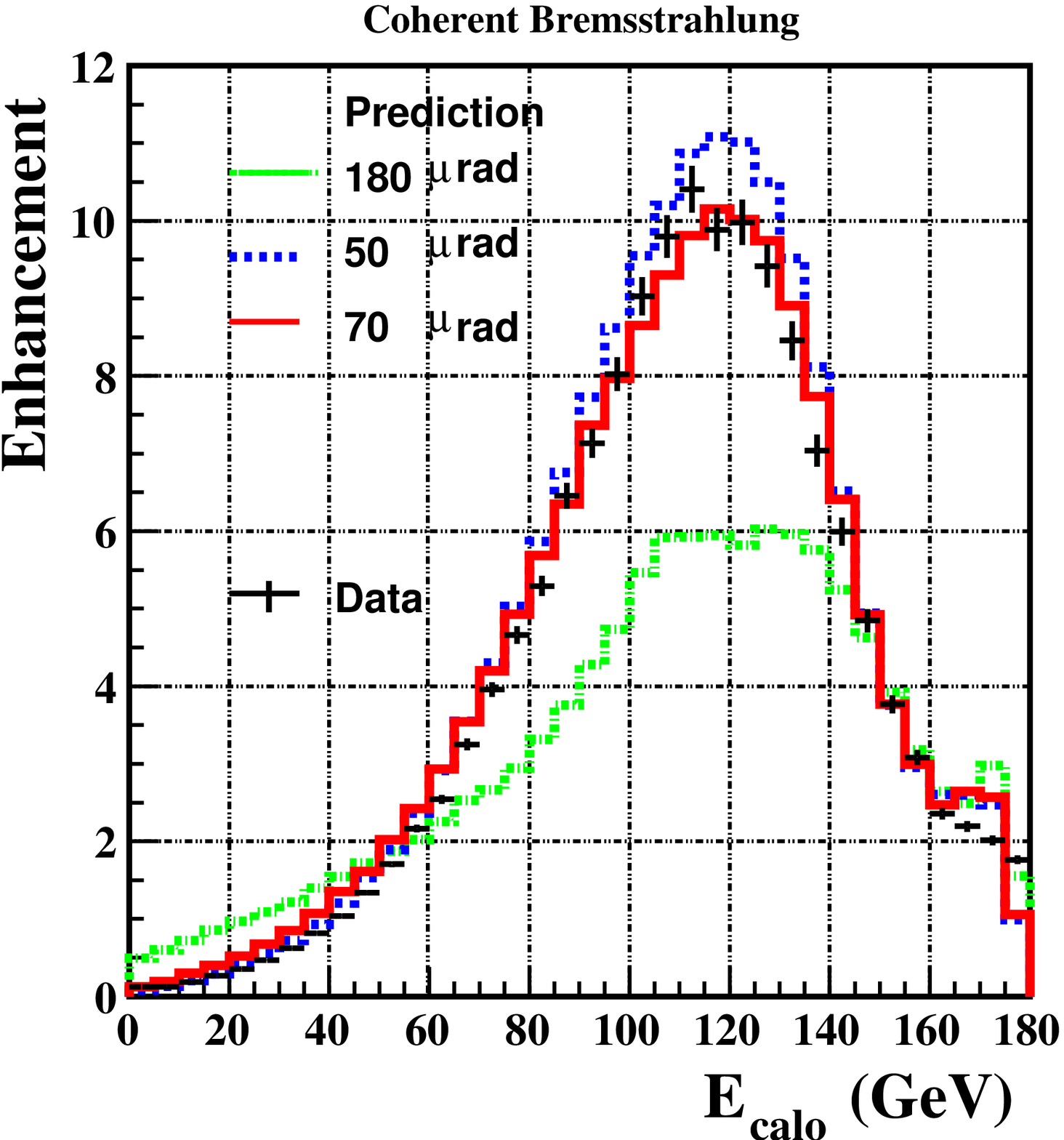}
\caption{\label{fig:enhancement} Enhancement of CB radiation data 
compared to  MC predictions.}
\end{figure}

An independent method of verifying the CB settings is looking at the total
electromagnetic radiation from the radiator crystal. Fig.~\ref{fig:pileup}
shows the total energy radiated ($EdN/dE$) measured by the calorimeter for
the radiator crystal aligned (bottom) and unaligned (top). In terms of the
radiation intensity spectrum, an unaligned crystal is identical to an
amorphous material. This radiation is called the Incoherent Bremsstrahlung
(IB) and it can be approximated by the familiar Bethe-Heitler
formula~\cite{BH}.

The increase in the CB radiation intensity spectrum is usually reported
with respect to the IB spectrum. This ratio, called the ``enhancement'',
is presented in Fig.~\ref{fig:enhancement} together with MC prediction for
CB angle at 70\,$\mu$rad. The agreement of the data with the enhancement
prediction is remarkable. The offline analysis could therefore be used to
monitor the angular settings of the radiator in time steps, to ensure the
crystal angular settings did not drift during the measurement.

\section{Results and Conclusions}

\begin{figure}[h]
\includegraphics[scale=0.478]{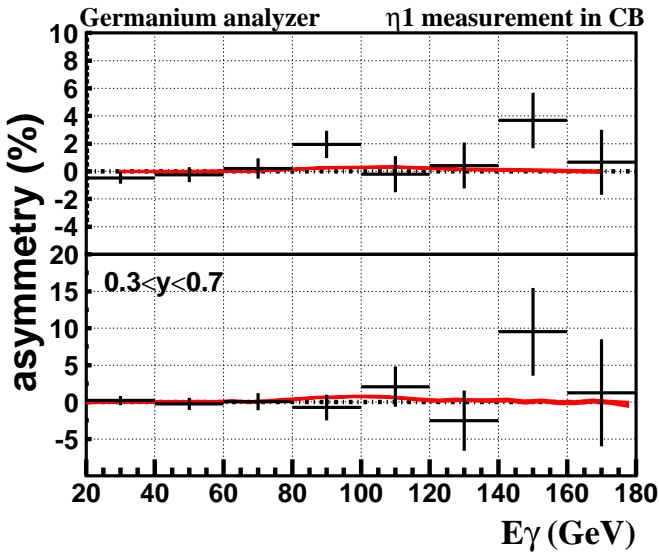}
\caption{\label{fig:eta1-rslt-ge} Asymmetry to determined $\eta_1$ 
component of the photon polarization with the Ge analyzer. The data at 
roll angles $\pi/4 + 5\pi/4$ are compared to $3\pi/4 + 7\pi/4$ {\it 
without (TOP)} and {\it with (BOTTOM)} the quasi-symmetrical pair 
selection.}
\end{figure}

Establishing the CB orientation allows the comparison of the predicted and
measured asymmetries for both linear polarisation components: $\eta_1$ and
$\eta_3$. Using all events, as well as events passing the
quasi-symmetrical pairs selection criteria, we see that, as expected, the
asymmetry in Fig.~\ref{fig:eta1-rslt-ge} is consistent with zero yielding
a vanishingly small $\eta_1$ component of the polarization.

\begin{figure}[h]
\includegraphics[scale=0.477]{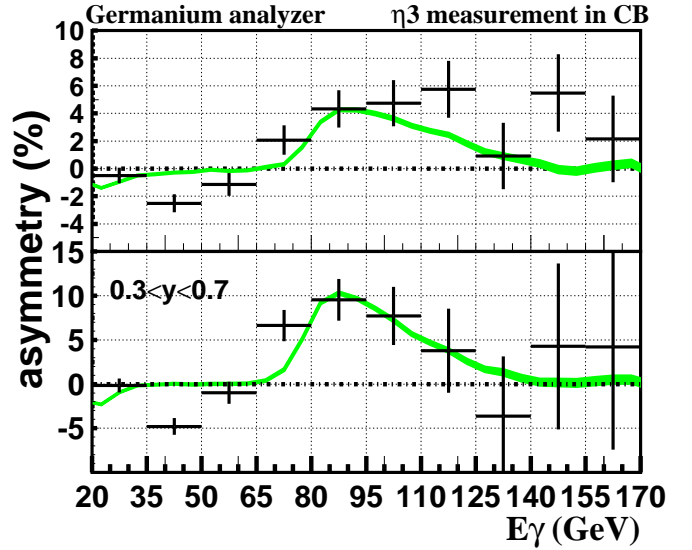}
\caption{\label{fig:eta3-rslt-ge} Asymmetry to determine the $\eta_3$ 
component of the photon polarization with the Ge analyzer. Measurements 
{\it without (TOP) } and {\it with (BOTTOM) } the quasi-symmetrical pair 
selection at roll angles $0 + \pi$ are compared to those at roll angles 
$\pi/2 + 3\pi/2$.}
\end{figure}

The measured asymmetry in the induced polarization direction ($\eta_3$)  
is presented in Fig.~\ref{fig:eta3-rslt-ge} without and with the $y$-cut
using the Ge analyzer crystal. The solid line represents the MC
predictions without any smearing in the spectrometer. The lower plot
represents the increase in the asymmetry due to quasi-symmetrical pairs
together with the statistical error associated with this increase. It thus
confirms the non statistical source of the asymmetry increase in the
70-110~GeV range. 

\begin{figure}[h]
\includegraphics[scale=0.47]{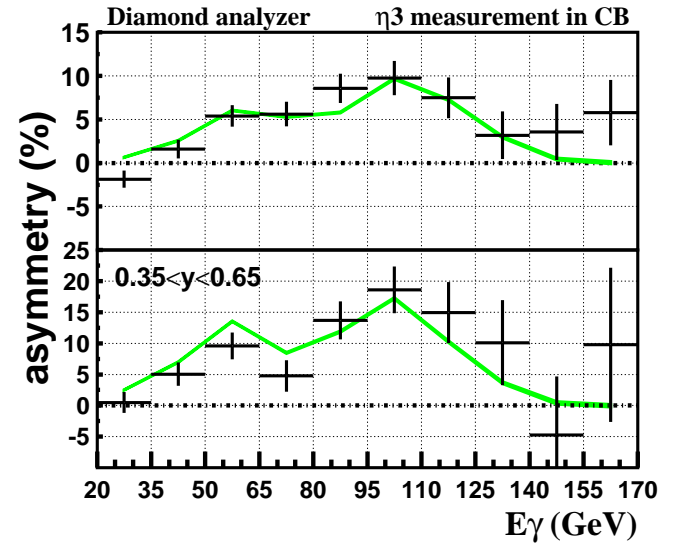}
\caption{\label{fig:eta3-rslt-di} Asymmetry measurements {\it without 
(TOP)} and {\it with (BOTTOM)} the quasi-symmetrical pair selection to 
determine $\eta_3$ component of the photon polarization with the {\it 
diamond} analyzer (Cf. Table~\ref{tab:data-sets}).}
\end{figure}

The same asymmetry as measured by the diamond analyzer is given in
Fig.~\ref{fig:eta3-rslt-di}. The top and middle plots are again the
asymmetry measurements as compared to the MC predictions without any
smearing, and the lower plot is the increase in the asymmetry due to the
$y$-cut. Comparing Fig.~\ref{fig:eta3-rslt-ge} and \ref{fig:eta3-rslt-di},
we conclude that the multi-tile synthetic diamond crystal is a better
choice than the Ge crystal as an analyzer, since for the same photon
polarization the former yields a larger asymmetry and thus enables a more
precise measurement. The diamond analyzer also allowed the measurement of
the photon polarization in the 30-70~GeV range, since it has some, albeit
small, analyzing power at these energies.

The theoretical predictions are based both on the calculation of the
energy dependent polarisation of photons produced by coherent
bremsstrahlung and the polarisation dependence of coherent pair
production, also as a function of incident energy. Thus the polarisation
sensitive versions of both coherent bremsstrahlung and coherent pair
production are needed together in the theoretical calculation that
predicts the measured asymmetry. The theoretical predictions are based
both on the calculation of the energy dependent polarisation of photons
produced by coherent bremsstrahlung and the polarisation dependance of
coherent pair production, also as a function of incident energy. The
agreement of this combined theory with the measured data is remarkable. It
is clear that, for the energy range of 30-170 GeV and the incident angle
phase space of this study, that the theory is sufficiently reliable and
well understood to support the development of applications of crystals as
polarimetry devices.  The calculation of the resolving power ($R$ in
equation~\ref{eq:asy-def}) is therefore reliable for the energy and angle
regimes discussed in the introduction.  The assymmetry measurements
therefore correspond to a measurement of the induced polarisation fo CB
for $\eta_3$ shown in Fig.~\ref{fig:pol-predict}. This has a maximum of
57\% at 70\,GeV. The experimental measured and predicted degree of linear
polarization (Stokes parameter $\eta_3$) are presented in~\cite{overview}.

These results then show the feasibility of the use of aligned crystals as
linearly polarized high energy photon beam sources. From the experimental
point of view, for the creation of a photon beam with a predictable
spectrum the crucial components are (i) high precision goniometers to
align the radiator crystal with respect to the electron beam and (ii) the
electron beam tracking chambers to monitor the angles of incidence on the
crystal surface. The predictability of the photon energy and polarization
is a good asset for designing future beamlines and experiments. These
results also establish the applicability of aligned crystals as
polarimeters for an accurate measurement of the photon polarization at
high energies. The important aspects are the analyzer material selection
and utilization of the quasi-symmetrical pairs. The use of synthetic
diamond as the analyzer crystal is found to be very promising due to its
availability, durability and high analyzing power.

The pair spectrometer enables the asymmetry measurement to be made for
single photons in multi photon events. If the photon multiplicity is low,
as it would be for laser generated beams with $E > 10$\,GeV, then a simple
multiplicity detector could be used to replace the more complex pair
spectrometer.  This is especially the case for a multiplicity detector
which is energy selective. Events with higher multiplicity are known to be
dominated by a single higher energy photon with the multiplicity
represented by lower energy photons.

Therefore, the crystal polarimetry technique developed here should also be
applicable in high energy photon beamlines as a fast monitoring tool. For
example, in a future $\gamma \gamma$ or $e \gamma$ collider the
quasi-online monitoring of the photon beam polarization could be achieved
using this crystal polarimetry method. In the most competitive designs of
such colliders~\cite{crapxing}, the photon beam after the interaction
region is always taken out to a beam dump, hence the destructive nature of
the crystal polarimetry technique does not constitute an impediment for
its utilization. This reported polarimetry technique was successfully used
in other studies of the NA59 research program~\cite{na59-l4,na59-sos}.

\begin{acknowledgments}

We dedicate this work to the memory of Friedel Sellschop. We express our
gratitude to CNRS, Grenoble for the crystal alignment and Messers DeBeers
Corporation for providing the high quality synthetic diamonds.  We are
grateful for the help and support of N. Doble, K. Elsener and H. Wahl. It
is a pleasure to thank the technical staff of the participating
laboratories and universities for their efforts in the construction and
operation of the experiment.

This research was partially supported by the Illinois Consortium for
Accelerator Research, agreement number~228-1001. UIU acknowledges support
from the Danish Natural Science research council, STENO grant no
J1-00-0568.

\end{acknowledgments}

\bibliography{na59-cb}

\end{document}